\documentstyle[aps,prb,floats]{revtex}
\def\be{\begin{equation}}
\def\ee{\end{equation}}
\def\bea{\begin{eqnarray}}
\def\eea{\end{eqnarray}}
\begin{document}
\draft
\title{ Modified Korteweg--de Vries Hierachies in Multiple--Times 
Variables and the Solutions of Modified Boussinesq Equations}
\author{M. A. Manna and V. Merle\\
Physique Math\'ematique et Th\'eorique,\\
CNRS - Universit\'e de Montpellier II
34095 Montpellier Cedex 05 (France)}
\maketitle
\begin{abstract}

We study solitary--wave and kink--wave solutions of a modified
Boussinesq equation through a multiple--time
reductive perturbation method. We use appropriated modified Korteweg--de
Vries hierarchies to eliminate secular producing terms in each order
of the perturbative scheme. We show that the multiple--time variables
needed to obtain a regular perturbative series are completely determined by
the associated linear theory in the case of a solitary--wave solution, 
but requires the knowledge of each order of the perturbative series in
the case of a kink--wave solution. These appropriate  multiple--time variables
allow us to show that the solitary--wave as well as the kink--wave solutions of the
modified Botussinesq equation are actually respectively a solitary--wave and a 
kink--wave satisfying all the equations of suitable modified
Korteweg--de Vries hierarchies.

\end{abstract}

\section{Introduction}
In this paper we study two model equations of the modified Boussinesq type
\be
u_{tt} - u_{xx} + u_{xxxx} +2\alpha(u^p)_{xx}=0\, ,
\label{1}
\ee
with $u(x,t)$ a one--dimensional field, $\alpha=\pm1$, $p=3$ and subscripts
denoting partial differentiation. These equations approximately describe
the propagation of waves in certain nonlinear dispersive systems. For
example they appear when we study the continuous limit of a
Fermi--Pasta--Ulam dynamical system with cubic nonlinearity (Dodd
{\em et al.} 1982).
They can
appear also governing the evolution of long internal waves of moderate
amplitude, (Ablowitz {\em et al.} 1973) or describing the dynamic of a
stretched string (Ablowitz {\em et al.} 1973; Mott 1973 ).

They are considered as intermediate long--wave equations since they
represent an intermediate dynamic, in complexity and completeness,  
situated between the complete dynamic of the full initial equations
describing any wave number and any amplitude, and
some strong long--waves and small- amplitude limits. Their
appropriated
long--wave and small--amplitude limit 
with a further restriction to unidirectional propagation yields 
the modified Korteweg--de Vries equation (mKdV).

Asymptotic methods are very often employed and very useful to study
problems of this type.
Hence, to study the evolution of long--waves of equations (\ref{1}), we will consider a perturbative
scheme based on the reductive perturbation method of Taniuti (Taniuti 1974).
We will include the classical slow space--variable $\xi$, but we will modify
 it by introducing an infinite number of slow time--variables: $\tau_3$,  
$\tau_5$, $\tau_7$, etc. As we are going to see, by using these slow
variables together with all the equations of an appropriated modified Korteweg--de Vries
hierarchy (Chern \& Peng 1979 ) , we can obtain the solitary--wave solution of the
modified Boussinesq equation (\ref{1}) with $\alpha=1$ (for shorteness mBI)
\be
u(x,t)=k\ {\rm sech} \left\{k \left(x-\sqrt{1-k^2}\right) t \right\},
\label{2}
\ee
as well as the kink--wave solution of the modified Boussinesq equation (\ref{1}) with $ \alpha=-1$ (mBII)
\be
u(x,t)=k\ {\rm tanh} \left\{k \left(x-\sqrt{1+2k^2}\right) t \right\}.
\label{3}
\ee
The solutions (\ref{2}) and (\ref{3}), written in the laboratory
coordinates $(x,y)$, are actually built up respectively from the solitary--wave
solution and the kink--wave solution of the whole set of equations of
appropriated mKdV hierarchies written in the slow space--variable $\xi$ and in
each one of the slow time--variables $\tau_3$, $\tau_5$, $\tau_7$, ...

These results follow from the general long--wave perturbation theory
and from the requirement of uniformity for large time of the associated
perturbative series. This last fact makes the perturbative series
truncates for solutions of type (\ref{2}) or (\ref{3}), rendering thus
exact solutions for the modified Boussinesq equations.
Furthermore, we will show that the elimination of the secular producing
terms in the perturbative series linked to the solution (\ref{2}), is
completely accounted for only the linear theory associated to (\ref{1}), which allows
us to know {\it "a priori" } all the constants which define the slow
times--variables $\tau^{'s}$ in function of $t$. These properly
normalized slow time coordinates automatically give us a perturbative
series which is free of secularities (uniform expansion).

On the other hand, to solve the same problem for the solution (\ref{3}),
that is, to eliminate the secular producing terms linked to solution
(\ref{3}), 
requires the knowledge of each term of the perturbative series.

This paper is organized as follows. In Sec.II the multiple time
formalism is introduced for the modified Boussinesq equations (\ref{1}), 
the first few evolution equations are obtained and the
problems associated with them are exhibited. In Sec.III we discuss how the mKdV
hierarchy shows up. In Sec.IV we show how they can be used, in the case
$\alpha=1$, to eliminate the soliton related secularities of the
evolution equations for the higher--order terms of the wave fields. In
Sec.V  by returning from the slow variables to the laboratory
coordinates we obtain the above mentioned relation between the
solitary--wave of mBI and the corresponding modified Korteweg--de Vries
hierarchy. Sec.VI is consecrated to the case of mBII $(\alpha=-1)$. In Sec.VII
we give a general proof that the symmetries of time derivatives lead to
the modified Korteweg--de Vries hierarchy. Finally in Sec.VIII we
summarize and discuss the results obtained.
 \section{The Multiple Time Formalisme: Mixed--Secular Terms}
In order to study the far--field dynamics of long--wave solutions of eq.(\ref{1}), we will
 need to define slow space and time variables. A small parameter
$\epsilon$ giving the long--wave character of the studied solution is
introduced via the definition
\be 
k=\epsilon \kappa \, ,
\label{4}
\ee
where $k$ is the wave number and $ \kappa $ a parameter of order
one. Accordingly, we define a slow space variable
\be
\xi=\epsilon (x-t) \, ,
\label{5}
\ee
as well as an infinity of slow time coordinates
\be
 \tau_3=\epsilon^3 t,\,\,\,\tau_5=\epsilon^5 t,\,\,\,\tau_7=\epsilon^7 t,....
\label{6}
\ee
Consequently, we have
\be
\frac{\partial}{\partial x}=\epsilon\frac{\partial}{\partial \xi}\,,
\label{7}
\ee
and
\be
\frac{\partial}{\partial t}=-\epsilon\frac{\partial}{\partial \xi}+
\epsilon^3\frac{\partial}{\partial
\tau_3}+\epsilon^5\frac{\partial}{\partial
\tau_5}+\epsilon^7\frac{\partial}{\partial \tau_7}+\,....
\label{8}
\ee
Moreover, we consider a small--amplitude solution of (\ref{1}) and we make the expansion
\be
u=\epsilon {\hat u} = \epsilon (u_0 + \epsilon^{2} u_{2} + \epsilon^{4}
u_{4} +\,...).
\label{9}
\ee
We suppose that $u_{2n}=u_{2n}(\xi,\tau_3,\tau_5,...)$ for
$n=0,1,2,...$ which corresponds to an extension of the function $u$ in the
 Sandri's  sense (Sandri 1965). Substituting eqs (\ref{7}) (\ref{8})
and (\ref{9}) into the Boussinesq equations (\ref{1}) the resulting
expression, up to terms of order $\epsilon^4$, is:

$$
\left [-2 \frac{\partial^2}{ \partial \xi \partial \tau_{3}} +
  \frac{\partial^4}{\partial \xi^4}+ \epsilon^2 \left  ( \frac{\partial^2}
 {\partial \tau_{3}^2}-2 \frac{\partial^2}{ \partial \xi \partial \tau_
{5}} \right) +\epsilon^4 \left ( -2 \frac{\partial^2}{ \partial \xi 
\partial \tau_{7}}+  2 \frac{\partial^2}{ \partial \tau_{3} \partial
 \tau_{5}} \right) + \, ...\right]{\hat u}
$$
\be
 2\alpha \frac{\partial^2}
{ \partial \xi^2}\left [u_{0}^{3} + \epsilon^{2}( 3u_{0}^{2}u_{2}) + 
\epsilon^{4}( 3u_{0}^{2}u_{4} + 3u_{0}u_{2}^{2}) +\, ...\right]=0\,,
\label{10}
\ee
At order $\epsilon^{0}$, after an integration in $\xi$, we get
\be
-2\frac{ \partial u_{0} }{ \partial \tau_{3} } + 
\frac{\partial^{3} u_{0}}
{\partial \xi^{3}} + 6\alpha u_{0}^{2}\frac{\partial u_{0}}
{\partial \xi} = 0,
\label{11}
\ee
which is the mKdV equation.

 At order $\epsilon^2$,
eq.(\ref{10}) yields, using (\ref{11}) and integrating once in $\xi$

$$
-2\frac{ \partial u_{2} }{ \partial \tau_{3} } + 
\frac{\partial^{3} u_{2}}
{\partial \xi^{3}} + 6\alpha \frac{\partial (u_{0}^2u_{2})}{\partial
\xi}=2\frac{\partial u_{0}}{\partial \tau_{5}} -
\frac{1}{4}\frac{\partial^{5} u_{0}}{\partial \xi^{5}}
-3\alpha \Big(\frac{\partial u_{0}}{\partial \xi}\Big)^3 - 9 \alpha u_{0}
 \frac{\partial u_{0}}{\partial \xi} \frac{\partial^2 u_{0}}{\partial
 \xi^2}
$$
\be
 -3\alpha u_{0}^2 \frac{\partial^3 u_{0}}{\partial
 \xi^3} - 9\alpha^2 u_{0}^4 \frac{\partial u_{0}}{\partial \xi}.
\label{12}
\ee
Equation (\ref{12}) is a linearized inhomogeneous mKdV whose general
solution consists of a sum of a general solution to the homogeneous equation
 and a particular solution to the nonhomogeneous equation.
 As it stands, (\ref{12}), presents two problems. First, the inhomogenity
(source term) is unknown because the evolution of $u_{0}$ in the time
$\tau_{5}$ is not known. The second problem is related to nonuniformity
of the expansion for $u$. When we considerer a soliton type solution of 
eq.(\ref{11}), case $\alpha = 1$, the term $\frac{\partial^5
u_{0}}{\partial \xi^5}$ is proportional to $\frac{\partial
u_{0}}{\partial \xi}$ which is a solution of the associated homogeneous
equation. Hence the general solution $u_{2}$ of (\ref{12})  
contains a term proportional to $\tau_{3}$ $\frac{\partial
u_{0}}{\partial \xi}$ which gives rise to a nonuniformity in the
perturbative series (mixed--secular term).

For $mBII$, case $\alpha = -1$, we have kink--type solutions to eq.
(\ref{11}). Solutions of type (\ref{2}) exist but are complex and we
will not consider them.
 In this case the linear term of eq.(\ref{12}), and actually some nonlinear
ones, produces secularity as well.

In the next sections we will deal with these two problems.  
\section{THE RISE OF THE MODIFIED KORTEWEG--DE VRIES HIERARCHY}
As we have seen, the field $u_{0}$ satisfies the mKdV equation in the
time $\tau_{3}$. The evolution of the same field $u_{0}$ in any of the
higher order times $\tau_{2n+1}$ can be obtained in the following way
 (Kraenkel {\em et al.} 1995).

  First, to have a well ordered perturbative scheme we impose that each
one of the equations for $u_{0,\tau_{2n+1}}$ be $ \epsilon $--independent
when passing from the slow variables ($u_{0}$ ,$\xi$, $\tau_{2n+1}$) to the
laboratory coordinates ($u$,$x$,$t$). This step selects all possible terms which can
appear in $u_{0,\tau_{2n+1}}$. For instance, the evolution of $u_{0}$ in
$\tau_{5}$ is restricted to the form 
\bea
u_{0 , \tau_ 5} & = & a u_{0 , 5\xi} + b u_0^2 u_{0 , 3\xi} + c u_0 u_{0 , \xi
} u_{0 , 2\xi} + d (u_{0 , \xi})^3 + e u_{0}^4 u_{0 , \xi} + f u_{0}
u_{0 , 4\xi} \nonumber \\ 
&+& g u_{0}^3 u_{0 , 2\xi} + h u_{0 , \xi} u_{0 , 3\xi} + i u_{0}^6 + j
(u_{0, 2\xi})^3, 
\label{13}
\eea
where a, b, c, d, e, f, g, h, i and j are unknown constants. Then, by imposing the
natural (in the multiple time formalism) compatibility condition
\be
\Big (u_{0 , \tau_{3}}\Big)_{\tau_{2n + 1}}= \Big (u_{0 , \tau_{2n +1}}
\Big )_{\tau_{3}}\,,
\label{14} 
\ee
with  $u_{0}$ satisfying mKdV in $\tau_{3}$, it is possible to determine
any $u_{0 , \tau_{2n + 1}}$ i.e to determine all the contants appearing in 
$u_{0 , \tau_{2n + 1}}$.

As it will be shown in Sec.VI, the resulting equations are those of the
mKdV hierarchy. In particular, for $u_{0 , \tau_{5}}$ and 
$u_{0 , \tau_{7}}$, using the mKdV in its canonical form (for
shorteness mKdVI) with $\alpha = 1$
\be
 u_{0, \tau_{3} } + 
 u_{0, 3\xi} + 6 u_{0}^{2}u_{0, \xi} = 0,
\label{15}
\ee
we obtain
\be
u_{0 , \tau_ 5}  =  \alpha_{5} \Big( u_{0 , 5\xi} + 10 u_0^2 u_{0 , 3\xi}
 + 40 u_{0} u_{0 , \xi} u_{0 , 2\xi} + 10 (u_{0 , \xi})^3 + 30 u_{0}^4 u_{0 ,
\xi} \Big)\,, 
\label{16}
\ee
and
\bea
u_{0 , \tau_{7}}  = - \alpha_{7} \Big(&+&u_{0 , 7\xi} + 84 u_{0} u_{0 , \xi}
u_{0 , 4\xi} + 560 u_{0}^3 u_{0 , \xi} u_{0 , 2\xi} + 14 u_{0}^2 u_{0 , 5\xi}
 + 140 u_{0} u_{0 , 2\xi} u_{0 , 3\xi} \nonumber\\
&+& 126 u_{0 , 3\xi} (u_{0 ,\xi})^2 + 182 u_{0 , \xi} (u_{0 , 2\xi})^2 + 
70 (u_{0})^4 u_{0 , 3\xi} + 420 (u_{0})^2 (u_{0 , \xi})^3 \nonumber\\
&+&140 u_{0 , \xi} (u_{0})^6 \Big)\,.
\label{17}
\eea
For the alternative form, mKdVII ($\alpha = -1$) 
\be
 u_{0, \tau_{3} } + 
 u_{0, 3\xi} - 6 u_{0}^{2}u_{0, \xi} = 0\,,
\label{18}
\ee
$u_{0 , \tau_{5}}$ and $u_{0 , \tau_{7}}$ are given by
\be
u_{0 , \tau_ 5}  =  \beta_{5} \Big( u_{0 , 5\xi} - 10 u_0^2 u_{0 , 3\xi}
 - 40 u_{0} u_{0 , \xi} u_{0 , 2\xi} - 10 (u_{0 , \xi})^3 + 30 u_{0}^4 u_{0 ,
\xi} \Big)\,, 
\label{19}
\ee
\bea
u_{0 , \tau_{7}}  =  \beta_{7} \Big(&-& u_{0 , 7\xi} + 84 u_{0} u_{0 , \xi}
u_{0 , 4\xi} - 560 u_{0}^3 u_{0 , \xi} u_{0 , 2\xi} + 14 u_{0}^2 u_{0 , 5\xi}
 + 140 u_{0} u_{0 , 2\xi} u_{0 , 3\xi} \nonumber\\
&+& 126 u_{0 , 3\xi} (u_{0 ,\xi})^2 + 182 u_{0 , \xi} (u_{0 , 2\xi})^2 - 
70 (u_{0})^4 u_{0 , 3\xi} - 420 (u_{0})^2 (u_{0 , \xi})^3 \nonumber\\
&+& 140 u_{0 , \xi} (u_{0})^6 \Big)\,.
\label{20}
\eea
The coefficients $\alpha_{5}$, $\alpha_{7}$,  $\beta_{5}$,  $\beta_{7}$, 
are free parameters which are not determined by the algebraic system originated
from eq.(\ref{14}). These free parameters are related to different
possible normalizations of the slow time variables and we will use them
to obtain a secular free perturbative solution of (\ref{1}).

We will see in Sec.V
and VI that the genesis of the values of these parameters are
rather different for the perturbation theory associated with the two cases 
$\alpha = 1$ or $\alpha = -1$.
 \section{Higher Order Evolution Equations For Modified Boussinesq I}
Let us consider the perturbation theory asociated with mBI ($\alpha =
1$ in eq.(\ref{1})) with $\tau_{5}$, $\tau_{7}$, ..., as defined by
(\ref{6}) and $\tau_{3}$ redefined by
\be
\tau_{3}= - \frac{ \epsilon^3 t}{2}.
\label{21}
\ee
Hence the operator (\ref{8}) now reads
\be
\frac{\partial}{\partial t}=-\epsilon\frac{\partial}{\partial \xi}-
\frac{1}{2}\epsilon^3\frac{\partial}{\partial
\tau_3}+\epsilon^5\frac{\partial}{\partial
\tau_5}+\epsilon^7\frac{\partial}{\partial \tau_7}+\,...
\label{22}
\ee 
At first order, this definition of $\tau_{3}$ gives the mKdVI in
canonical form (\ref{15}). At order $\epsilon^2$ we obtain the following
equation:
\be
{\cal L_{I}}( u_{2}) = 2 u_{0 , \tau_{5}}
-\frac{1}{4}\int_{ -\infty }^\xi {u_{0 , 2\tau_{3}
} d\xi^{'}},
\label{23}
\ee
where $\cal L_{I}$ is the linearized mKdVI operator defined by
\be 
{\cal L_{I}}(v)= v_{\tau_{3}} + v_{3\xi} + 6 (u_{0}^2 v)_{\xi}.   
\label{24}
\ee
Substituting $u_{0 , \tau_{3}}$ and  $u_{0 , \tau_{5}}$ from (\ref{15}) and
(\ref{16})  we have
\bea
{\cal L_{I}}( u_{2}) = &2& \Big( \alpha_{5} - \frac{1}{8} \Big) u_{0 , 5\xi} +  
\Big( 20\alpha_{5} - 3 \Big) u_{0}^2 u_{0 , 3\xi} + \Big(80\alpha_{5} - 9
\Big) u_{0} u_{0 , \xi}  u_{0 , 2\xi}  \nonumber\\ & & +\Big(20\alpha_{5} - 3\Big)
 (u_{0 , \xi})^3 +  \Big(60\alpha_{5} - 9 \Big) u_{0}^4 u_{0 , \xi}.
\label{25}
\eea 
If we assume the solution of the mKdVI (eq.(\ref{15})) to be the
solitary--wave solution 
\be
u_{0} = \kappa\ {\rm sech}\ ( \kappa \xi - \kappa^3 \tau_{3} + \theta ),
\label{26}
\ee
where $\theta$ is a phase, it is easy to see that only the term $u_{0
, 5\xi}$ is a secular--producing term. It can be eliminated if we choose
$\alpha_{5} = \frac{1}{8}$. In this case, eq.(\ref{25}) becomes 
\be
{\cal L_{I}}( u_{2}) = -\frac{1}{2} u_{0}^2 u_{0 , 3\xi} +  u_{0} u_{0 , 
\xi}u_{0 , 2\xi} - \frac{1}{2}(u_{0, \xi})^3 - \frac{3}{2} u_{0}^4 u_{0 ,
\xi}. 
\label{27}
\ee
Using for $u_{0}$ the solitary--wave solution (\ref{26}), we see that the
right--hand side of eq.(\ref{27}) vanishes, leading to 
\be
{\cal L_{I}} (u_{2}) = 0,
\label{28}
\ee
which is a homogeneous linearized mKdVI equation, for which we will
consider the trivial solution
\be
u_{2} = 0.
\label{29}
\ee
At order $\epsilon^4$, and already assuming that $u_{2} = 0$, we obtain
\be
{\cal L_{I}}( u_{4}) = 2 u_{0 , \tau_{7}}
 + \int_{ -\infty }^\xi {u_{0 , \tau_{3}
, \tau_{5}} d\xi^{'}}.
\label{30}
\ee
Using equations (\ref{15}) and  (\ref{16}), with $\alpha_{5}=\frac{1}{8}$,
 to express respectively $u_{0 ,
 \tau_{3}}$ and  $u_{0 , \tau_{5}}$ we obtain
\bea
{\cal L_{I}}( u_{4}) = &2&\Big( u_{0 , \tau_{7}} - \frac{1}{16} u_{0 , 
7\xi}\Big) - \frac{65}{4} u_{0 , 3\xi} (u_{0 , \xi})^{2} - \frac{35}{2}
 u_{0} u_{0 , 2\xi}u_{0 , 3\xi} - 10 u_{0} u_{0 , \xi} u_{0 , 4\xi}
\nonumber\\ 
&-2& u_{0 , 5\xi} u_{0}^{2} - \frac{45}{2} u_{0 , \xi} (u_{0 ,
2\xi})^{2} - 
\frac{105}{2} u_{0}^2 (u_{0 , \xi})^{2} - 75 u_{0 , \xi} u_{0 , 2\xi}
u_{0}^3 - \frac{45}{4} u_{0 , 3\xi} u_{0}^4  \nonumber\\
&-\frac{45}{2}& u_{0 , \xi} (u_{0})^6\,.
\label{31}
\eea
Now the source term proportional to $u_{0 , 7\xi}$ is the only resonant.
Then, in the same way we did before, we first use the higher mKdVI, 
eq.(\ref{17}), to express  $u_{0 , \tau_{7}}$ and then we choose the
free parameter $\alpha_{7}$ in such a way to eliminate the resonant term
$u_{0 , 7\xi}$. This choice corresponds to $\alpha_{7} =
-\frac{1}{16}$, which brings eq.(\ref{31}) to the form
\bea
{\cal L_{I}}( u_{4}) = &\frac{1}{2}& u_{0} u_{0 , \xi} u_{0 , 4\xi}- 5 
 u_{0 , \xi} u_{0 , 2\xi} u_{0}^3 - \frac{1}{4} u_{0 , 5\xi} u_{0}^2 - 
\frac{1}{2} u_{0 , 3\xi} (u_{0 , \xi})^{2} \nonumber\\
&+\frac{1}{4}& u_{0 , \xi} (u_{0 , 2\xi})^{2} - \frac{5}{2} u_{0 , 3\xi}
 u_{0}^{4} - 5 u_{0 , \xi} u_{0}^{6}\,. 
\label{32}
\eea 
Substituting again the solitary--wave solution (\ref{26}) for $u_{0}$, 
we see that the nonhomogeneous term of eq.(\ref{32}) vanishes,
leading to 
\be
{\cal L_{I}}( u_{4}) = 0.
\label{33}
\ee
Again, we take the trivial solution
\be
u_{4} = 0.
\label{34}
\ee
This is a general result that will repeat at any higher order. For
$n \geq 1$ the evolution equation for $u_{2n}$, after using the mKdVI
hierarchy equations to express $u_{0 , \tau_{2n + 1}}$ and after eliminating
the secular producing term coming from the solitary--wave solution (\ref{26}), is
given by an homogeneous linearized mKdVI equation.

Consequently, the solution $u_{2n} = 0$, for $n \geq 1$, can be assumed
for any higher order. 
\section{Back to the laboratory coordinates. Connection to the
dispersion relation}
Actually, to obtain a perturbative scheme free of secular producing terms at
 any
higher order, we assume that $u_{0}$ is the solitary--wave solution of all the
equations of the mKdVI hierarchy, each one in a different slow--time
variable. Such a solution may be obtained and is given by
\be
u_{0}( \xi, \tau_{3}, \tau_{5}, \tau_{7},...) = \kappa \;{\rm sech} \Big[
\kappa \xi - \kappa^3 \tau_{3} + \kappa^5 \alpha_{5} \tau_{5} - \kappa^7
 \alpha_{7} \tau_{7} + ...\Big].
\label{35}
\ee
First, recall that we have expanded $ u $ according to eq.(\ref{9}).
Thereafter, we have found a particular solution in which $u_{2n} = 0$
for $n \geq 1$. Consequently expansion (\ref{9}) truncates, leading to an
exact solution of the form 
\be
u=\epsilon u_{0}\,,
\label{36}
\ee
with $u_{0}$ given by eq.(\ref{35}). Moreover, the slow variables
($\kappa$, $\xi$, $\tau_{2n+1}$) are related to the laboratory ones ( $k$, $x$,
$t$ ), respectively by eqs. (\ref{4}), (\ref{5}), (\ref{6}) and (\ref{21})
and we have found later that
\be
\alpha_{5}=\frac{1}{8} ,\:\:\:\:\:\: \alpha_{7}=-\frac{1}{16},\:....
\label{37}
\ee
Then, in the laboratory coordinates, the exact solution (\ref{36}) is
written  
\be
u( k, x, t ) = k\; {\rm sech} k\Big[ x - \Big(1 - \frac{1}{2}k^2 - \frac
{1}{8}k^4 - \frac{1}{16}k^6 - ...,\Big) t \Big].
\label{38}
\ee
Now, the series appearing inside the parenthesis can be summed:
\be
 1 - \frac{1}{2}k^2 - \frac{1}{8}k^4 - \frac{1}{16}k^6 - ... = \sqrt{
1 - k^2 }.
\label{39}
\ee 
Consequently, we get
\be
u = k\; {\rm sech}\; k \;\Big( x - \sqrt{1 - k^2}\; t \Big),
\label{40}
\ee
which is the solitary--wave solution of mBI.

 Let us now take the linear
mBI dispersion relation
\be
\omega (k) = k\;\sqrt{ 1 + k^2 }\,.
\label{41}
\ee
Its long--wave expansion $(k = \epsilon \kappa )$ is given by
\be
\omega (k) = \epsilon \kappa + \frac{1}{2}\epsilon^3 \kappa^3 - \frac{1}
{8}\epsilon^5 \kappa^5 + \frac{1}{16}\epsilon^7 \kappa^7 + ....
\label{42}
\ee
In passing we notice that the absolute value of the coefficients of the expansion
coincide exactly with those found: $\alpha_{3}$ to obtain the mKdVI in
canonical form and $\alpha_{5}$,  $\alpha_{7}$, ..., necessary to
eliminate the secular producing term in each order of the perturbative scheme.

With this expansion, the solution of the associated linear mBI reads
\be
u = exp\; i\Big[\kappa \epsilon \Big( x - t \Big) - \frac{1}{2}\epsilon^3 
\kappa^3 t + \frac{1}{8}\epsilon^5 \kappa^5 t - \frac{1}{16}\epsilon^7 
\kappa^7 t + ...\Big].
\label{43}
\ee
Therefore, if we define from the begining, as given by this
expression, the properly normalized slow time coordinates
\be
\tau_{3} = -\frac{1}{2}\epsilon^3 t,\;\;\; \tau_{5} =  \frac{1}{8}\epsilon^5 t, 
\;\;\; \tau_{7} =  -\frac{1}{16}\epsilon^7 t,\;\;...,
\label{44}
\ee
the mKdVI (\ref{15}) will be obtained at first order and the resulting
perturbative theory will be automatically free of secular producing terms
for the soliton solution.
\section{HIGHER ORDER EVOLUTION EQUATION FOR THE MODIFIED\\ BOUSSINESQ II}
We consider now the case $\alpha = -1$ in eq.(\ref{1}). We assume the same
operators $\frac{\partial }{\partial x}$ and $\frac{\partial }
{\partial t}$ given by (\ref{7}) and (\ref{22}), and the same expansion
for $u$. We obtain at order $\epsilon^{0}$ and 
$\epsilon^{2}$ the following equations
\be
 u_{0, \tau_{3} } + 
 u_{0, 3\xi} - 6 u_{0}^{2}u_{0, \xi} = 0,
\label{45}
\ee 
which is mKdVII, and 
\be
{\cal L_{II}}( u_{2}) = 2 u_{0 , \tau_{5}}
-\frac{1}{4}\int_{ -\infty }^\xi {u_{0 , 2\tau_{3}
} d\xi^{'}},
\label{46}
\ee 
where ${\cal L_{II}}$ is the linearized mKdVII operator. Using eq.(\ref{45})
to express $u_{0, \tau_{3}}$ and eq.(\ref{19}) to express $u_{0, \tau_{5}}$ in
(\ref{46}) we obtain
\bea
{\cal L_{II}}( u_{2}) = & & \Big( 2\beta_{5} - \frac{1}{4} \Big) u_{0 , 5\xi} -
\Big( 20\beta_{5} - 3 \Big) u_{0}^2 u_{0 , 3\xi} - \Big(80\beta_{5} - 9
\Big) u_{0} u_{0 , \xi}  u_{0 , 2\xi}  \nonumber\\ & &
-\Big(20\beta_{5} - 3\Big)
 (u_{0 , \xi})^3 + \Big(60\beta_{5} - 9 \Big) u_{0}^4 u_{0 , \xi}.
\label{47}
\eea
In this case the real solution of mKdVII is of kink--wave type. It reads
\be
u_{0}(x,t)=\kappa\; {\rm tanh} \left(\kappa \xi + 2\kappa^3\tau_{3} + \theta 
 \right).
\label{48}
\ee
When we substitute (\ref{48}) in (\ref{47}), resonant terms appear. Contrary to
the case of mBI, they do not come only from the linear term, but also from the
nonlinear ones. They are proportional to
\be
 u_{0 , \xi} = \kappa\;{\rm sech}^2
\Big( \kappa \xi + 2\kappa^3 \tau_{3} + \theta \Big)
\label{49},
\ee
and we have
\be
 {\cal L_{II}}( u_{2}) = -\kappa\;{\rm sech}^2\Big( \kappa \xi + 2\kappa^3
 \tau_{3} + \theta \Big)\Big(1 - 12\beta_{5}\Big).
\label{50}
\ee
Hence, the value
\be
\beta_{5} = \frac{1}{12}
\label{51}
\ee
eliminates the secular producing term at order $\epsilon^2$, and gives ${\cal L_{II}}( u_{2}) =
0$. We then assume the trivial solution $u_{2} = 0$.
At order $\epsilon^4$ we have 
\be
{\cal L_{II}}( u_{4}) = 2 u_{0 , \tau_{7}}
 + \int_{ -\infty }^\xi {u_{0 , \tau_{3}
, \tau_{5}} d\xi^{'}}.
\label{52}
\ee
Using (\ref{18}) (\ref{19}) and (\ref{20}) to express respectively $u_{0 , \tau_{3}}$, 
 $u_{0 , \tau_{5}}$ and  $u_{0 , \tau_{7}}$ we obtain, (using (\ref{48}))
\be
 {\cal L_{II}}( u_{4}) = \kappa\;{\rm sech}^2\Big( \kappa \xi + 2\kappa^3
 \tau_{3} + \theta \Big)\Big(1 + 40\beta_{7}\Big).
\label{53}
\ee
We choose
\be
\beta_{7} = -\frac{1}{40}
\label{54}
\ee
to eliminate the secular producing term at this order. Again, we take the
trivial solution
\be
u_{4} = 0.
\label{55}
\ee
Actually, as in the previous case, we can assume $u_{2n}= 0$ for $n\geq 1$ and we have the
exact solution
\be
u = \epsilon^2 u_{0},
\label{56}
\ee
where $ u_{0}$ is the kink--wave solution to all equations of  the mKdVII
hierarchy, which reads
\be
u = \kappa\, {\rm tahn}\, \Big(\kappa \xi + 2\kappa^3 \tau_{3} +  6\kappa^5\beta_{5}
 \tau_{5} +  20 \kappa^7 \beta _{7}\tau_{7} + ...\Big).
\label{57}
\ee
In the laboratory coordinates, the exact solution (\ref{57}) is written
as
\be
 u = k\, {\rm tahn}\, k\Big[x - \Big(1 + k^2 - \frac{1}{2}k^4 + \frac{1}
{2}k^6 - ...\Big)\,t \Big].
\label{58}
\ee
The series appearing inside the parenthesis can be summed:
\be
1 + k^2 - \frac{1}{2}k^4 + \frac{1}{2}k^6 - ... = \sqrt{1 + 2k^2 },
\label{59}
\ee
and we obtain the kink--wave solution of mBII
\be
u = k\, {\rm tanh}\,\Big(kx - k\sqrt{1 + 2k^2}\,t\Big).
\label{60}
\ee  
To obtain (\ref{60}) we used the definitions
\be
\tau_{3} = -\frac{1}{2}\epsilon^3t,\,\,\,\, \tau_{5} =
\frac{1}{12}\epsilon^5t,\,\,\,\,\tau_{7} =
-\frac{1}{40}\epsilon^7t,\,
\,\,\,...\,\,.
\label{61}
\ee 
In this case the cofficients are not those obtained from the long--wave
expansion of the linear dispersion relation of mBII.
\section{Commutativity of time derivatives implies the mKdV
hierarchy: A general proof}
In this section we give a general proof that the symmetries of time
derivatives, with the scale invariance requirement, leads to the mKdV
hierarchy in $\tau_{5}$, $\tau_{7}$,..., for a field which satisfies mKdV in $
\tau_{3}$. We do this for mKdVI.

Let $M_{n}$ be a polynomial in terms of the form
\be
u^{a_{0}}u^{a_{1}}_{\xi}u^{a_{2}}_{2\xi}...u^{a_{l}}_{l\xi},
\label{62}
\ee 
with $a_{i} \in  N$. We define the Rank (R) of $M_{n}$ as
\be
R(M_{n}) = \sum_{j = 0}^{l} (1 + j)a_{j}.
\label{63}
\ee
Let us now consider our principle which requires that each
one of the evolutions equations
\be
u_{\tau_{3}},\,\,\,\,u_{\tau_{5}},\,\,\,\,u_{\tau_{7}}, ...,u_{\tau_
{2n - 3}} \mbox{\null\hspace{3cm}} n = 2, \, 3, \, ...,
\label{64}
\ee
be $\epsilon$-- independent when passing from the slow $(u_{0}, \xi,
\tau_{2n - 1})$ to the laboratory coordinates $(u, x, t)$. It is easy to
see that this is equivalent to require that
\be
u_{\tau_{2n - 3}} = M_{n}, \mbox{\null\hspace{3cm}}  n =
2,\,\,3,\,\,...,
\label{65}
\ee
with
\be
 R(M_{n}) = 2n - 2, \mbox{\null\hspace{3cm}} n = 2,\,\,3,\,\,....
\label{66}
\ee

Equation (\ref{65}) and the condition (\ref{66}) give us the
correct terms in the expression of $u_{\tau_{2n - 3}}$ but they do not
determine their coefficients.

Let us recall that the Maurer--Cartan equation associated with the group of
 $2 \times 2$ real unimodular matrices (Chern \& Peng 1979), leads to the mKdV
hierarchy -- in variables $x$, $t$ -- in the form
\be
u_{t} = u^{-1} R_{n + 1, x}\,,
\label{67}
\ee
where
\be
R_{0} = -1.
\label{68}
\ee
$R_{n}$ for $n \geq 1$ may be generated from the recursion
formula
\be
u^{-1} R_{n + 1,x} = \frac{1}{4} \Big(u^{-1}R_{n,x}\Big)_{2x} 
+\Big(u R_{n}\Big)_{x}.
\label{69}
\ee

Let us renormalize the coefficients of highest order
derivative term in each one of the equations of the hierarchy
 to $+1$ or $-1$ by defining
\be
A_{n,x} = (-1)^{n}4^{n - 1}u^{-1}R_{n,x}\,.
\label{70}
\ee 
Hence, we write the higher--order mKdVI equations
\be
u_{t} = A_{n - 1,x}\,\,\,\,  \mbox{\null\hspace{3cm}} n = 1,\,\,2,\,\,...,
\label{71}
\ee
with
\be  
A_{n + 1,x} = -A_{n,3x} -
4\Big(u\int_{-\infty}^{x}uA_{n,\xi}\,d\xi\Big) + u_{x}\delta_{n,0}\,,
\mbox{\null\hspace{2cm}} n = 0,\,\,1,\,\,...,
\label{72}
\ee
where $\delta_{n,0}$ is the Kronecker symbol. With the above statements we
 establish the following theorem\\
${\rm THEOREM}$\\
If $\Big(u_{\tau_{3}}\Big)_{\tau_{2n - 3}} =
\Big(u_{\tau_{2n - 3}}\Big)_{\tau_{3}}$ with the invariance condition
$u_{\tau_{2n - 3}} = M_{n}$ where 
\be
M_{3} = - u_{3\xi} - 6u^2u_{\xi}\,,
\label{73}
\ee
 and
\be
R(M_{n}) = 2n - 2\,,
\label{74}
\ee
then we have
\be
M_{n} = A_{n - 1,\xi}\,, \mbox{\null\hspace{3cm}} n = 1,\,\,2,\,\,..., 
\label{75}
\ee
where the $A_{n}$ are the $n^{th}$ conserved densites of mKdVI.\\
Proof:\\
The commutativity of time derivatives and the invariance condition lead
to the linearized mKdVI for $M_{n}$
\be
M_{n,\tau_{3}} + M_{n,3\xi} + 6\Big(u^2 M_{n}\Big)_{\xi} = 0\,.
\label{76}
\ee
Thus we will prove that
\be
\Big(A_{n - 1,\xi}\Big)_{\tau_{3}} + \Big(A_{n - 1,\xi}\Big)_{3\xi} +  
6\Big(u^2 A_{n - 1,\xi}\Big)_{\xi}  = 0\,.
\label{77}
\ee
We proceed by induction. For $n = 3 $, eq.(\ref{77}) gives
 mKdVI, and (\ref{75}) clearly holds if we assume
that the arbitrary function of integration in $\tau_{3}$ that appears
is zero. This is justified because each $M_{n}$ must be a  polynomial in
terms of (\ref{62}).
 Assuming that it holds for $n - 2$, we have
\be
\Big(A_{n-2,\xi}\Big)_{\tau_{3}} + \Big(A_{n-2,\xi}\Big)_{3\xi} +  
6\Big(u^2 A_{n-2,\xi}\Big)_{\xi}  = 0\,.
\label{78}
\ee
We use (\ref{72}), the mKdVI and the inductive hypothesis (\ref{78})
to show -- after some heavy algebra -- that 
\be
\Big(A_{n - 1,\xi}\Big)_{\tau_{3}} + \Big(A_{n - 1,\xi}\Big)_{3\xi} +  
6\Big(u^2 A_{n - 1,\xi}\Big)_{\xi}  = 0\,.
\label{79}
\ee
This proves the theorem.

\section{Conclusion}
We have applied a multiple--time version of the reductive perturbation
method to study the solitary--wave type solution and the kink--wave
type solution of two families of modified Boussinesq model equations.

In the first case we have eliminated the solitary--wave related secular
producing terms through the use of the equations of an appropiated
modified Korteweg--de Vries hierarchy. We have shown that the
solitary--wave solution of the associated modified Boussinesq equation
is given by a solitary--wave satisfying, in the slow variables, all the
equations of the modified Korteweg--de Vries hierarchy. Accordingly,
while the modified Korteweg--de Vries solitary--wave only depends on
one slow variable $\tau_{3}$, the solitary--wave solution of the
modified
Boussinesq equation can be thought of as depending on the infinite slow
time variables.

 Hence, solitary--wave solutions of intermediate model
equations, like the modified Boussinesq, contain  complete information
concerning all degrees of long--waves present in the system.

In the second case, we have eliminated the kink--wave related secular
producing terms through the use of another appropriated modified
Korteweg--de Vries hierarchy. We have shown that the kink--wave solution of
the associated modified Boussinesq equation can be built from the
kink--wave solution of all equations of the corresponding modified
Korteweg--de Vries hierarchy.

The main difference between the
solitary--wave case and the kink--wave case is situated in
the formulae of tranformation giving the slow--time variables
$\tau^{,s}$ in function of the laboratory time coordinate $t$. In the
first case a formula exists and is given by
the long--wave expansion of the linear dispersion relation of the
initial equation, independently of the associated perturbation theory.

In the second case such a formula does not exist and we know the 
$\tau^{,s}$ in funtion of $t$ only {\it"a posteriori"}, that is, by inspecting
each order of the perturbative theory.

In the solitary--wave case the sources of secularities at each
order of the associated perturbative series are the linear terms
only. If we linearize the perturbative series ($u_{n} = 0$ for $n =
1,2,....$) we obtain the equations of the associated Fourier transform 
of the extended initial function (Sandri 1965) if, and only if, the
definitions of $\tau^{,s}$ like a function of $t$ was done according to
the development of the linear dispersion relation for long--waves.

Hence, the requirement of a perturbative scheme free of solitary--wave related 
secularities and the
existence of a compatible linear limit of the theory are completely equivalent.

In the kink--wave case, there are two sources of secularity at each order: 
the linear terms and also some nonlinear ones. Hence the
elimination of the linear terms, or the compatibility with the
associated Fourier theory is not sufficient to obtain a regular
perturbative series.
We must carry out a second renormalization of the
kink--wave frequency which gives us the right coefficients transforming
$\tau_{2n + 1}$ in a function of $t$. In this paper (Sec.VI) we realized these two
renormalizations in only one step, but they can be realized separately.
In this case the second renormalization of the frequency clearly appears
like a reminiscence of the celebrated Stokes' hypothesis
 (Whitham 1974) on the frequency amplitude--dependence in water waves.

Let us remember that obstacles to asymptotic integrability can be appear in 
the case of nonintegrable systems (Kodama \& Mikhailov 1995). Such obstacles
was exibited in the case of the nonintegrable RLW equation (Kraenkel {\it
et al} 1996). For integrable systems, like the one consider here, the multiple scale
method
will be able to handle both, the solitary--wave and the $N$--soliton
related secularities since no obstacles to asymptotic integrability
will be present.x

Finaly in Sec.VII we have shown that the commutativity of time derivatives,
together with the scale invariance requirement, lead to the mKdV
hierarchy in $\tau_{5}$, $\tau_{7}$,..., for a field which satisfies mKdV in $
\tau_{3}$. 

\vspace {1 cm}
\section*{Acknowledgements}
The authors would like to thank J. L\'eon, R.A. Kraenkel and J.G. Pereira
for useful discussions.

\end{document}